\documentclass[conference]{IEEEtran}
\usepackage{color,graphicx}
\usepackage{url}
\usepackage{comment}
\usepackage{amsmath}
\usepackage{algorithmic}
\usepackage{algorithm}
\usepackage{epstopdf}
\usepackage{epsfig}
\usepackage[T1]{fontenc}
\usepackage{txfonts}
\usepackage{cite}
\usepackage{array,multirow,graphicx}

\makeatletter

\newcommand{\Rmnum}[1]{\expandafter\@slowromancap\romannumeral #1@}
\makeatother
\begin{document}
\title{Enhancing Spectral Utilization by Maximizing the Reuse in LTE Network}
\author{\IEEEauthorblockN{Yuva Kumar S.$^*$, Vanlin Sathya$^\dag$, Sreenath Ramanath$^*$}
\IEEEauthorblockA{$^*$Lekha Wireless Solutions, Bangalore, India\\$^\dag$ University of Chicago, Illinois, USA.\\
{e-mail: $^*$[yuva.kumar, sreenath]@lekhawireless.com, $^\dag$vanlin@uchicago.edu}}}
\maketitle
\begin{abstract}
Need for increased spectral efficiency is key to improve quality of experience for next generation wireless applications like online gaming, HD Video, etc.,. In our work, we consider a LTE Device-to-device (D2D) network where LTE UEs have primary access to the spectrum and D2D pairs have secondary access. To enhance spectral efficiency, BS can offload the traffic by activating multiple D2D pairs within the serving cell. This ensures that the same radio resource will be reused across the primary LTE UEs and different D2D pairs. In this context, we propose to enable more D2D secondary users in the serving cell, by utilizing neighboring BS spectrum to fairly co-exist with neighboring LTE primary users. We model the system and show via extensive simulations, that the above configuration guarantees good throughput for the D2D pairs in the serving cell while ensuring that the primary LTE throughput demand is not compromised. 
\end{abstract}

\begin{IEEEkeywords}
LTE, Device-to-device (D2D), Cognitive Spectrum Access.
\end{IEEEkeywords}

\section{Introduction}
Due to heavy penetration of smartphone and tablets, we have seen a steady rise in the usage of the mobile applications (like YouTube, online gaming, Facebook, Netflix, WhatsApp, etc.) everyday by the cellular users. This situation forces the mobile/cellular operators to increase or reuse the available spectrum. Due to the high capital expenditure (CAPEX) on the spectrum used, reusing the existing (available) spectrum will be a good solution. In this context, one of the feasible solutions would be to deploy more small cells (like Femto or Pico) under a single Macro base station (BS), which will boost the overall spectrum efficiency with the increase of co-channel co-tier interference. An alternate solution to offload mobile traffic at the Macro BS and boost the spectral usage efficiency is to enable/activate more device to device (D2D) communication within the coverage of the BS. In LTE, D2D communications allow UEs to communicate directly without the intervention of eNodeB (eNB). D2D technology has been suggested as a method by the 3GPP for offloading the users traffic and increase the overall spectral efficiency. It also has been included in LTE release 12, as a guidance for implementing. 
\par D2D is one of the most promising technology towards 5G for improving resource utilization, enhancing UE throughput and extending battery lifetime. Due to its short-range communication, D2Ds have the potential to reuse the same radio resource with different transmission power across different D2D pairs. 
The challenges involved in enabling D2D communications are resource allocation, interference management, handovers, power control, session management, etc. There can be chances where the spectral reuse (in order to efficiently offload the BS traffic using D2D) might not be possible beyond a certain extent (which will result in more interference to the existing cellular users). To overcome this, it is also possible that one can intelligently allocate the neighboring BS spectrum and efficiently reuse the spectrum among D2Ds.
Efficient spectrum reuse and co-existence with existing technology is one of the key research topics. 
As the mobile operators demand for more spectrum, telecom regulatory bodies have been considering release of unused spectrum which has been used in applications such as radars, television, military applications, etc., to operators using cognitive radio technology. During the course of migration there is a need to come up with an alternate mechanism to manage the heavy traffic demands. 

The work presented in this paper looks into aspects of enhancing overall spectral efficiency. It is proposed that the operator will intelligently use the neighboring BS radio resources for assigning it to the additional D2D pairs. Here neighboring BS will be primary user and additional D2D pairs will be the secondary users. Primary user wouldn't be affected by secondary user transmission because of the relatively large path loss distance. This will boost overall system throughput.
 Hence, our work focuses on reusing the neighboring spectrum to small cell and D2D in such a way that the overall spectrum efficiency is maximized.
As operators are reusing the neighboring BS spectrum among small cells and D2Ds with shorter transmission power in the serving cell, it'll not interfere with the primary transmissions in the neighboring cell. 
Hence enabling the additional spectrum for small cell and D2D in a cognitive manner as discussed will tremendously increase the overall spectral reuse efficiency.
\par This paper is organized as follows. Section II describes the overview of the state of art in D2D communications. 
LTE and D2D co-existence is described in Section III. 
The system is modelled and analyzed in Section IV. 
Numerical and simulation results are provided in Section V and the conclusions are provided in Section VI.
\section{Related Work}
Telecom industry and research community works on boosting the spectrum efficiency by deploying more small cells and D2D under a single Macro BS. To attain a good SINR authors in~\cite{gc,el} deployed small cells inside the buildings by considering Macro-Femto interference. Also, authors proposed an efficient placement and dynamic power control SON (Self organizing Network) algorithm which optimally plan the small cells deployment and dynamically adjusts the transmission power of small cell based on the occupancy of Macro users in the interference zone terrain. To solve this, two Mixed Integer Programming (MIP) methods were proposed, namely: \textit{Minimize number of small cells method} which guarantees SINR threshold ($SINR_{th}$) -2dB for all indoor users and \textit{optimal small cell power allocation method} which guarantees $SINR_{th}$ (- 4 dB) for indoor users with the Macro users SINR degradation lesser than 2dB. 
The authors in \cite{lteu} consider a scenario having both Uplink (UL) and Downlink (DL) traffic for UEs and D2Ds with the assumption that they use different Resource Blocks (RBs) in a given Transmission Time Interval (TTI). 
They had proposed a semi distributed algorithm for allocation of the RBs.
The RBs are divided between D2Ds and UE in a centralized fashion. In which the transmission power for D2Ds are assigned in a distributed manner. Author did not use the same RB to UE and D2D links hence the overall reuse spectrum is not efficient.
While solving the RB allocation problem, the authors have assumed that the D2D links transmit at maximum power. 
Spectral reuse can be improved by controlling the transmission power. 
D2D transmission with dedicated resource and power control is considered in \cite{pad2d}.
\par Inter operator spectrum sharing is a well-studied problem in cellular networks \cite{ss,pimrc}. 
Unlike cellular spectrum sharing, this work is a first of its kind, which shows the benefits of co-existence of LTE and  D2D in a traditional cellular licensed band system. 
Fair co-existence among traditional cellular technologies like LTE and Wi-Fi is one of the key research areas. 
In this co-existence scenario, co-operative and non co-operative channel selection, scheduling \cite{sch}, mobility management \cite{handover} are key challenges.
In \cite{d2du} more D2D links are enabled by utilizing the unlicensed spectrum, appropriately designing LTE-Unlicensed (LTE-U) protocol (\emph{i.e.,} changing the listen before talk (LBT) mechanism). 
But traditionally the LTE-U LBT protocol is designed for collision avoidance techniques among UEs. 
Hence, fair co-existence among Wi-Fi, LTE-U and D2D is one of the key research challenges. 
In \cite{unly}, \cite{unly1} enabling D2D using underlay approach has been well studied.
In \cite{mobihoc,dyspan} authors discuss about efficient usage of unlicensed spectrum in a cognitive fashion. 
An analytical model for evaluating the baseline performance of the two different radio access technologies are provided.
 
In our work, we propose to efficiently reuse the neighboring cell spectrum to enable more D2D links to fairly co-exist with primary cellular users.
\section{Proposed Work}\label{SA}
In this section, we first describe the system model and then propose the cognitive algorithm for enhancement of spectral reuse in a given TTI. 
\subsection{System Model}
\begin{figure}
\begin{center} 
\includegraphics[width=9cm]{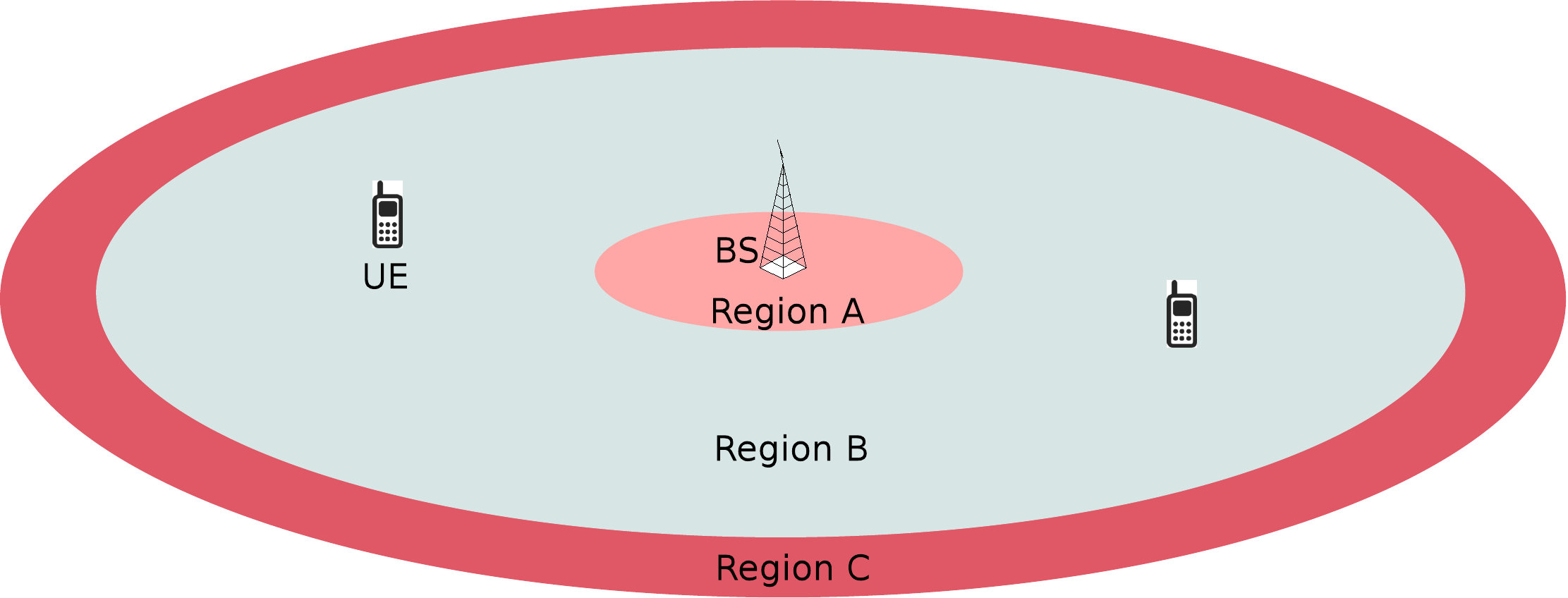}
\caption{Different Region inside a Base Station Cell}
\label{arch}
\end{center}
\end{figure}

In LTE, cellular geographical area of the BS can be divided in to three regions as shown in Fig. \ref{arch} and is explained as follows.
\begin{itemize}
\item \textbf{Region A:} This is defined as the field near to the BS and can be characterized with high power in the Downlink (DL).
\item \textbf{Region B:} This is defined as the field between region A and C, where the D2D links can work effectively without much interference to or from UE and BS.
\item \textbf{Region C:} This is defined as the field far from BS and can be characterized with high power in Uplink (UL) to reach BS from the UE.
\end{itemize}
\par
\begin{figure*}[htb!] 
\centering
\includegraphics[scale = .425]{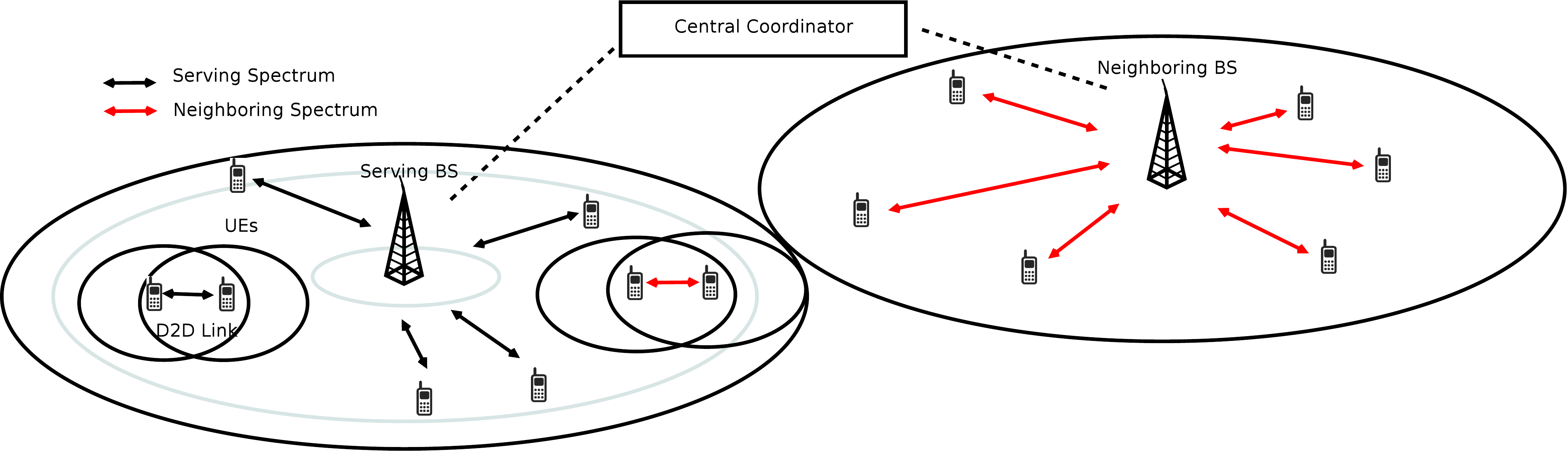}
\caption{D2D System Model with Cognitive Radio Access}
\label{sysmodel}
\end{figure*}

 We consider such an LTE network, with a BS serving a set $U$ of UEs as shown in Fig. \ref{sysmodel}. 
UEs will have both UL and DL demand. 
A set $L$ is the number of D2D links having spectral demand. 
The BS operates in hybrid mode (traditional cellular or cognitive) depending on the traffic load. 
Serving BS is centrally coordinated with neighboring BSs. 
We assume that a BS has the information such as the number of active users, traffic demand, location of the users (using position reference signal (PRS)), etc. 
One UE will be served at a given TTI consuming all the RBs.
We assume traffic at the BS is according to Poisson Point Process (PPP). 
When traffic at the BS increases, to overcome the resource crunch, D2D links are initiated in the serving cell by reusing the spectrum of the serving BS using cognitive underlay approach. 
Enabling D2D links with the spectrum of the serving BS will be challenging in Region A and Region C due to high DL and UL interference. 
Hence, in our work we enable D2Ds with the serving BS spectrum in Region B, where the serving BS interference impact will be less.
A utility function defines the number of allowed D2D links in the region B, while ensuring the quality of link between UE and BS in the serving cell. 
If the number of D2D links permitted by the utility function is less than the $L$, database based underlay CR-D2D links are initiated within the serving BS. 
CR-D2D links will be enabled using the available spectrum from the neighboring BSs in any region (\emph{i.e.,} Region A, B or C). 
Number of permissible CR-D2D links without affecting the link quality of the neighboring BSs and UEs is determined by the utility function.
\subsection{LTE - D2D Interference Modeling}
According to \cite{sgpp1}, interference in large wireless networks can be modelled according to a PPP. 
In a more generalized form, \cite{sgpp} models interference caused by low power nodes such as D2D and homogeneous interference caused by UE as a gamma distribution function. 
Let $X$ denote an exclusion region around the desired receiver with radius $R$. 
This region might be quite small, just a few meters in radius, and is designed to avoid the case where the low-power node is co-located with the target receiver. 
Let us consider that the interference is associated with a PPP given by  $\psi$ and the transmitting power of the D2D link is given by $P_d$, and the transmitter density by  $\lambda$. 
The received interference power at the UEs caused by the low power D2D nodes for a given TTI is given by,
\begin{equation}\label{eq1} 
I_d = \sum_{{d\in\psi} \textbackslash X} \frac{G_dL_dP_d}{l(R_d)}
\end{equation} 
For interferer $d$, large-scale fading $L_d$ follows a log-normal distribution with $\sigma$, while the small-scale fading distribution $G_d$ is $\Gamma[d_g , \theta_g]$. ${l(R_d)}$ is a non-singular path-loss model.
Similarly, interference caused by the UE to the D2D for a given TTI is given by:
\begin{equation}\label{eq2} 
I_u = \frac{G_uL_uP_u}{l(R_u)}
\end{equation} 
Where $G_u$ is the small-scale fading,  $L_u$ is the large-scale fading, $P_u$ is transmit power. ${l(R_u)}$ is a non-singular path-loss model.
\par Then, Signal to Noise plus Interference Ratio (SINR) of UE $u$ is given by:\\
\begin{equation}\label{eq3} 
SINR_{u} =\frac{G_{BS}L_{BS}P_{BS}}{N+I_d}
\end{equation}
where, $G_{BS}$ and  $L_{BS}$ are the channel fading from  BS to  UE, $P_{BS}$ is the transmission power of BS, $N$ is the overall noise in the system, $I_d$ is the interference caused by  all D2D links which are scheduled in the TTI of interest. \\
SINR of the receiver $d^{r}$ in the D2D link $d$ is given by:\\
\begin{equation}\label{eq4}
SINR_{d^r}=\frac{G_dL_dP_d}{N+I_u+I_{d/d^r}}
\end{equation}
where $G_{d}$ and  $L_{d}$ are the channel fading for the D2D link, $P_{d}$ is the transmission power of D2D, $N$ is the overall noise in the system,  $I_u$ is the interference caused by UE scheduled in the TTI of interest $I_{d/d^r}$ is the interference caused by the other D2D links which are scheduled in the TTI of interest. \\
\subsection{LTE Cognitive Radio D2D Algorithm}
LTE Cognitive Radio D2D (\textit{LTE-CR-D2D}) algorithm allocates maximum number of D2D links  which could be served in a particular TTI. \textit{LTE-CR-D2D} algorithm ensures throughput fairness to all D2D links. 
The system throughput is a measure of the system performance which changes based on the transmit power of the source and the interference in the system. For computing the system throughput in each TTI, we formulate an utility function as defined in Eqns. (\ref{eq6}) and (\ref{eq8}).
Average rate of a link is given by
\begin{equation}\label{eq5} 
\textnormal{\textbf E\{ln(1 + SINR)\}}
\end{equation}
Aim of this utility function is to maximize the logarithmic sum of the average rates of all the D2D links given as
\begin{equation}\label{eq6}
Maximize\displaystyle\sum\limits_{{r \in L}} log({\textbf E\{ln(1 + SINR_{d^r})\}})
\end{equation}
For a given bandwidth $B$ and spectral efficiency $\eta_{(SINR_{d^r})}$ corresponding to D2D $d^r$ with SINR  $SINR_{d^r}$, using LTE CQI table~\cite{cqi} average rate of a link can also be represented as,
\begin{equation}\label{eq7}
{R}_{d^r} = B\eta_{(SINR_{d^r})}
\end{equation}
Which gives utility function as
\begin{equation}\label{eq8}
Maximize\displaystyle\sum\limits_{{r \in L}} log({R}_{d^r})
\end{equation}
 \textit{LTE-CR-D2D} algorithm ensures that the maximum number of D2D links are being served in the serving cell. 
 It also ensures that the SINR of the UE doesn't fall below a permissible limit, which is defined as a function of $SINR_{th}$.
 
 \begin{algorithm}[htb!]
\caption{LTE-CR-D2D Algorithm}\label{Alg1}
\textbf{Input $:$ $PoissonArrival$, $SINR_{th}$ } \\
\textbf{Output $:$ $maxUtility$, $maxUtility_{cr}$, $D2D_{served}$, $CRD2D_{served}$}\\
\hrule
\begin{algorithmic} [1]
\STATE $L \leftarrow PoissonArrival$
\FOR{k =1 to $L$}
\STATE Calculate $SNR_{u}$ using path-loss model
\STATE Calculate serving cell $SINR_{u}$ using \eqref{eq3}
\STATE $maxUtility^{k} \leftarrow utility$
\IF{$SINR_{u}<SNR-SINR_{th}$}
\STATE break;
\ENDIF
\IF {($maxUtility^{k} < maxUtility^{k-1}$)}
\STATE break;
\ENDIF
\STATE $D2D_{served} \leftarrow k$
\ENDFOR
\IF {($D2D_{served}  < L$)}
\FOR{k = $D2D_{served}$+1 to $L$}
\STATE Calculate $SNR_{u}$ using path-loss model
\STATE Calculate neighboring cell $SINR_{u}$ using \eqref{eq3}
\STATE $maxUtility^{k}_{cr} \leftarrow utility$
\IF{$SINR_{u}<SNR-SINR_{th}$}
\STATE break;
\ENDIF
\IF {($maxUtility^{k}_{cr} < maxUtility^{k-1}$)}
\STATE break;
\ENDIF
\STATE $CRD2D_{served} \leftarrow k$
\ENDFOR
\ENDIF
\end{algorithmic}
\end{algorithm}
  
\begin{algorithm}[htb!]
\caption{Utility Function}
\textbf{Input $:$} $k$
\textbf{Output $:$} utility
\hrule
\begin{algorithmic}[1]
\FOR{r =1 to $k$}
\STATE Calculate $SINR_{d^r}$ using  \eqref{eq4}
\STATE Calculate ${R}_{d^r}$ using \eqref{eq7}
\ENDFOR
\STATE utility = $\displaystyle\sum\limits_{{r=1}}^{k} log({R}_{d^r})$;
\STATE return utility;
\end{algorithmic}
\end{algorithm}
Overview of the \textit{LTE-CR-D2D} algorithm is presented in Algorithm \ref{Alg1}. 
\section{Experimentation and Results}\label{exp}
The system model which is described in Section~\ref{SA} is simulated using MATLAB. 
We consider a system having two BSs which is of 30m height and the coverage of each BS is 500m. 
Other simulation parameters are as shown in Table \ref{table1}.
\begin{table}[htb!]
\caption{Simulation parameters}
\centering
\begin{tabular}{|p{4cm}| p{3.3cm}|}
\hline\bfseries
Parameter&\bfseries Value \\ [0.4ex]
\hline
$P_{d}^{max}$(Maximum D2D Tx power)& 0.1 W\\ 
\hline
$P_{BS}$(BS Tx power)& 39.8 W\\
\hline
Number of UEs in each BS & 50  \\
\hline
D2D Arrival Rate ($\lambda$) in serving BS & 20    \\
\hline
Bandwidth used in each BS & 10 MHZ (\emph{i.e.,} 50 RB) \\
\hline
UE deployment & Random \\
\hline
Traffic & Downlink \\
\hline
LTE Mode& FDD \\
\hline
$SINR_{th}$ & 0,2,4,6,8 dB\\
\hline
Path loss model for cellular link& $128.1 + 37.6log10(d[km])$  \\ 
\hline
Path loss model for D2D link& $148+40log10(d[km])$ \\
\hline
\end{tabular}
\label{table1}
\end{table} 
The D2D arrivals are modeled according to a Poisson Point Process with $\lambda$ as the arrival rate. 
$SINR_{th}$ is used as permissible threshold for UEs to ensure throughput of the UE is maintained.
Performance of the spectrum reuse is evaluated using the following parameters.
\begin{itemize}
\item Spectral Efficiency $\eta$, is defined as the number of D2D links which reused the spectrum per TTI.
\item Link Blocking Rate, is defined as the number of D2D links not getting served at a given TTI in the serving cell.
\item Spectrum Efficiency Gain is the attained throughput of the system by spectrum reuse. 
\end{itemize}
 
\begin{figure}
\centering
\includegraphics[scale=.36]{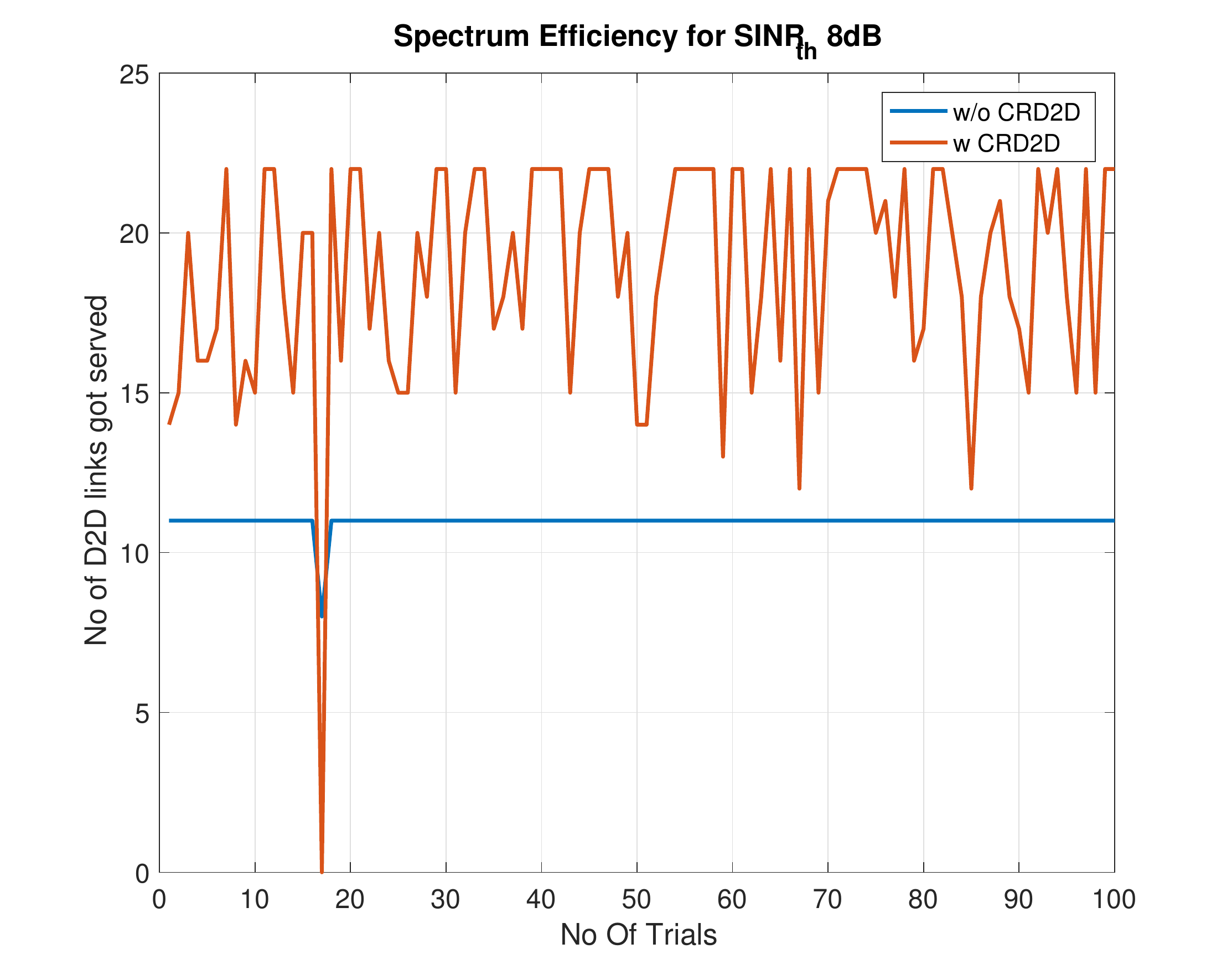}
 \caption{Number of D2D got served in the Serving BS for $SINR_{th}$  8dB}\label{se10}
\end{figure}
\begin{figure}
\centering
\includegraphics[scale=.36]{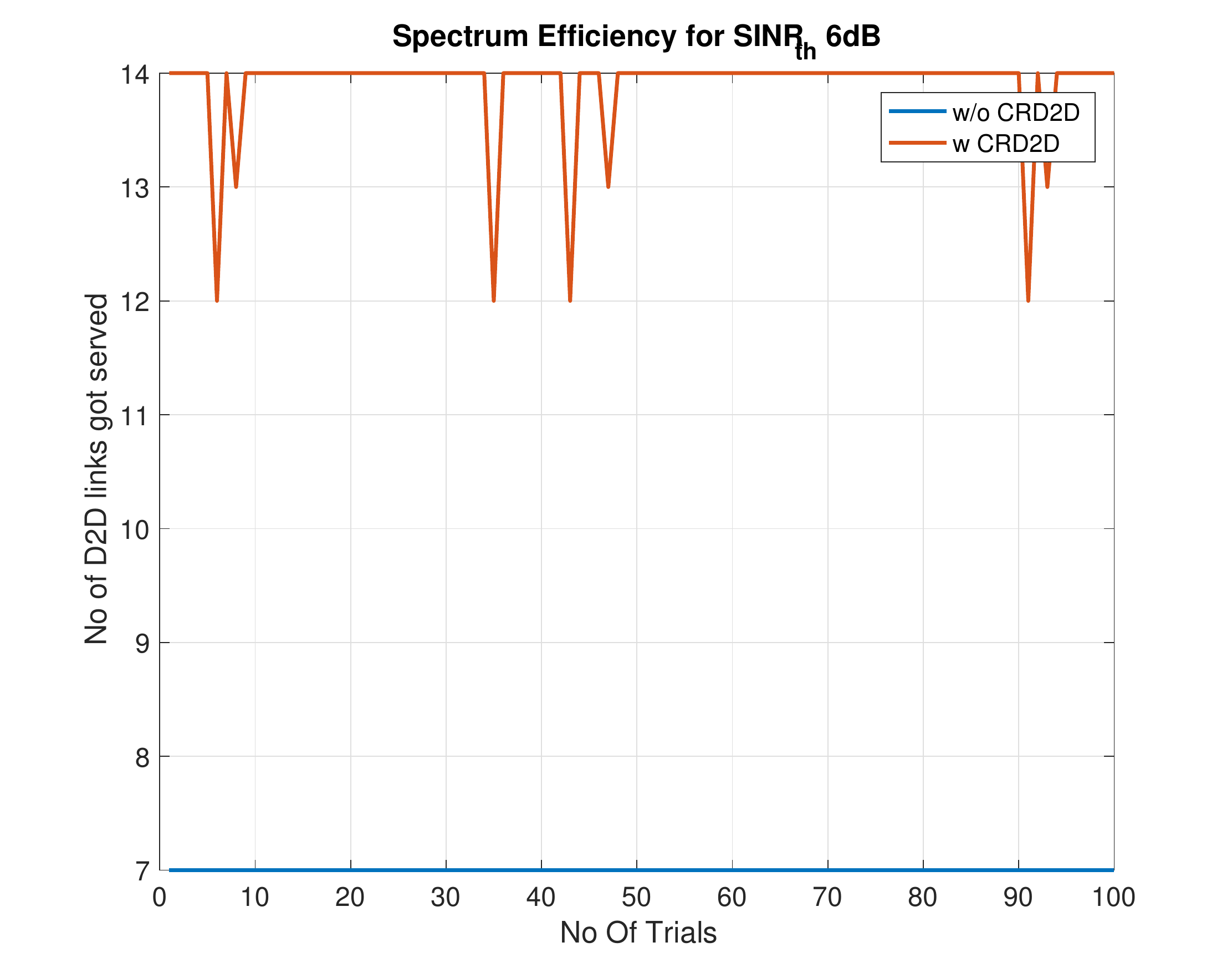}
 \caption{Number of D2D got served in the Serving BS for $SINR_{th}$  6dB}\label{se6}
\end{figure}
\begin{figure}
\includegraphics[scale=.36]{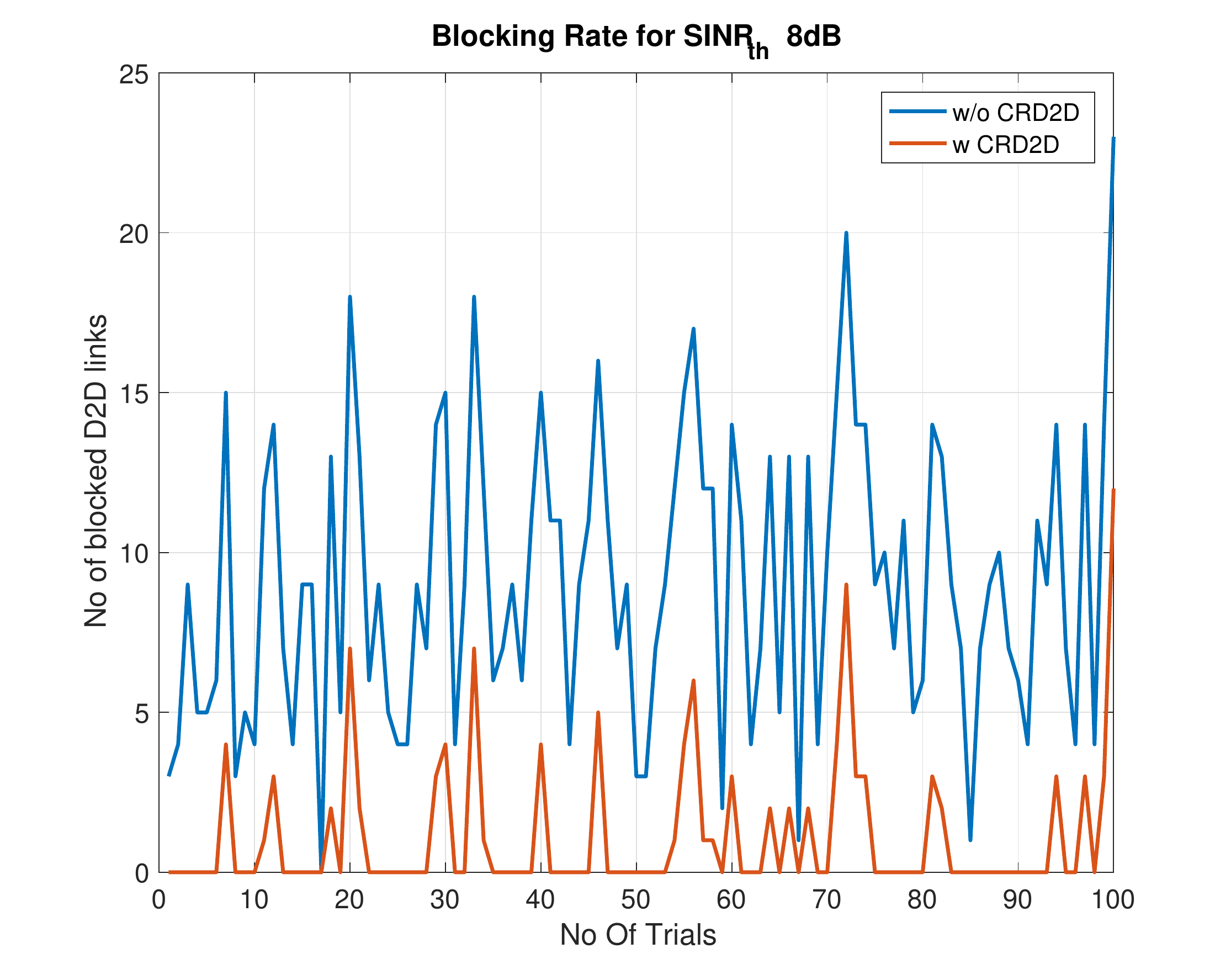}
 \caption{D2D Blocking Rate in the Serving BS  for $SINR_{th}$ 8dB}\label{br10}
\end{figure}
\begin{figure}
\includegraphics[scale=.36]{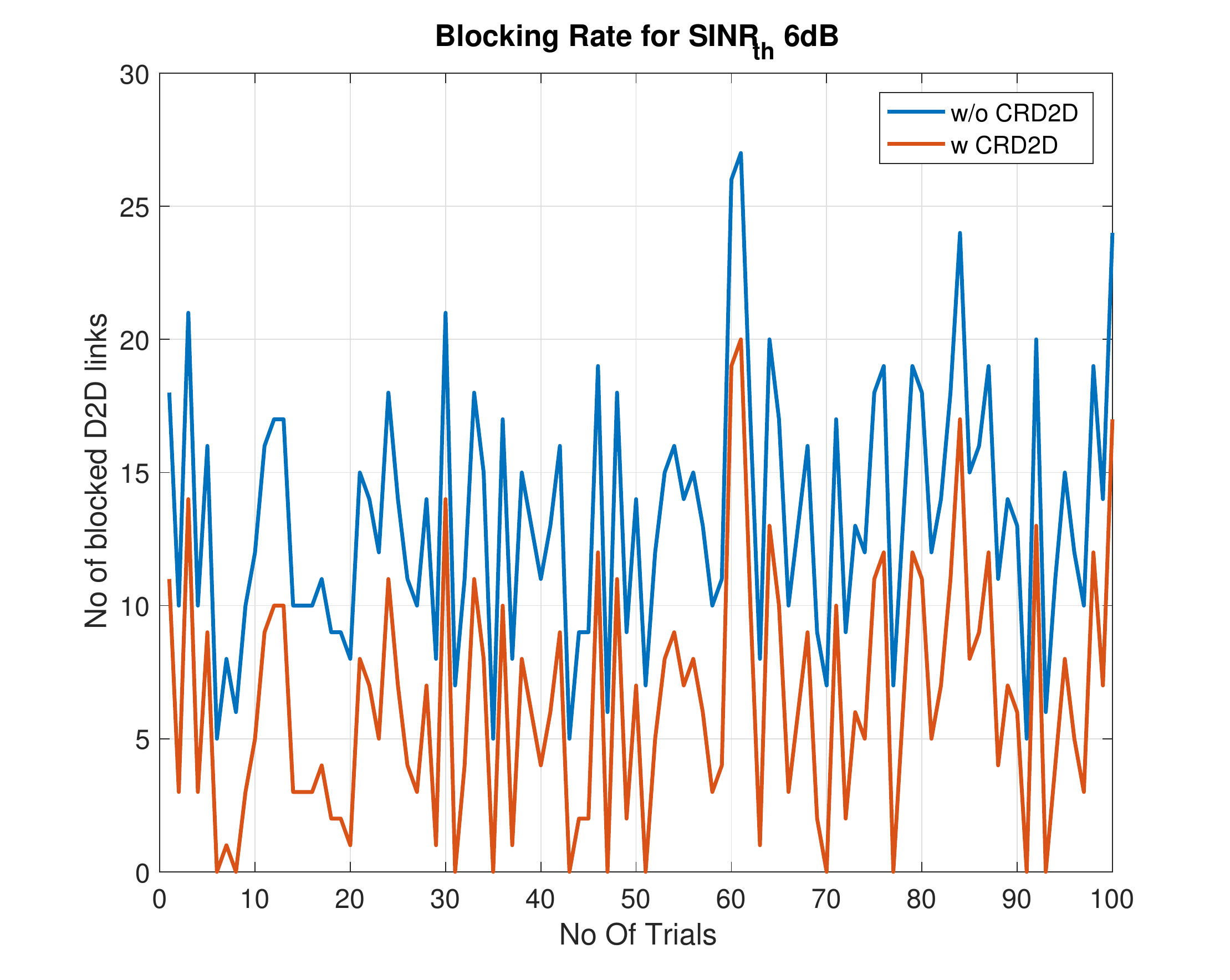}
 \caption{D2D Blocking Rate in the Serving BS  for $SINR_{th}$ 6dB}\label{br6}
\end{figure}
\begin{figure}
\includegraphics[scale=.36]{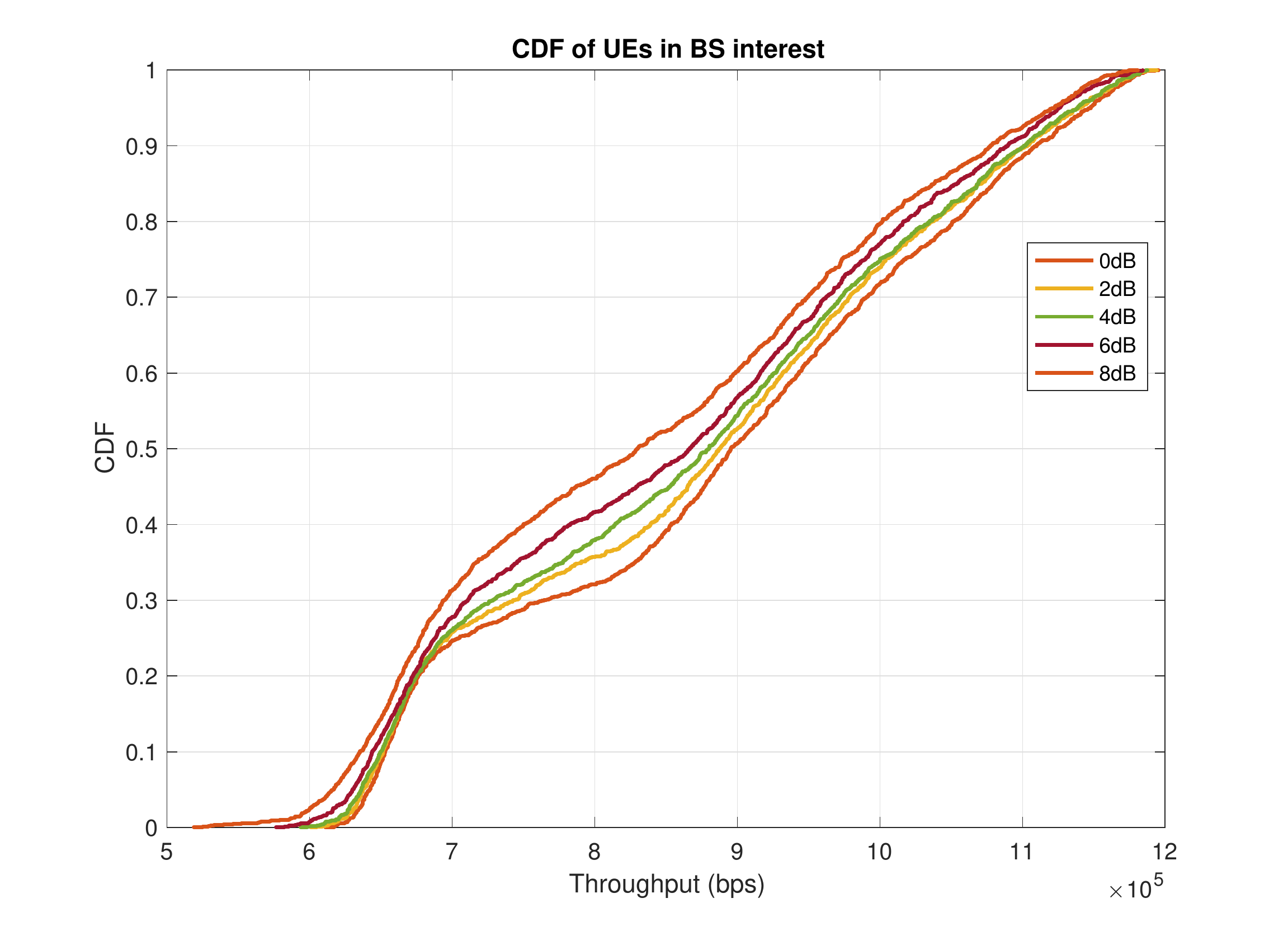}
 \caption{CDF of UEs throughput}\label{fp22}
\end{figure}
\begin{figure}
\includegraphics[scale=.36]{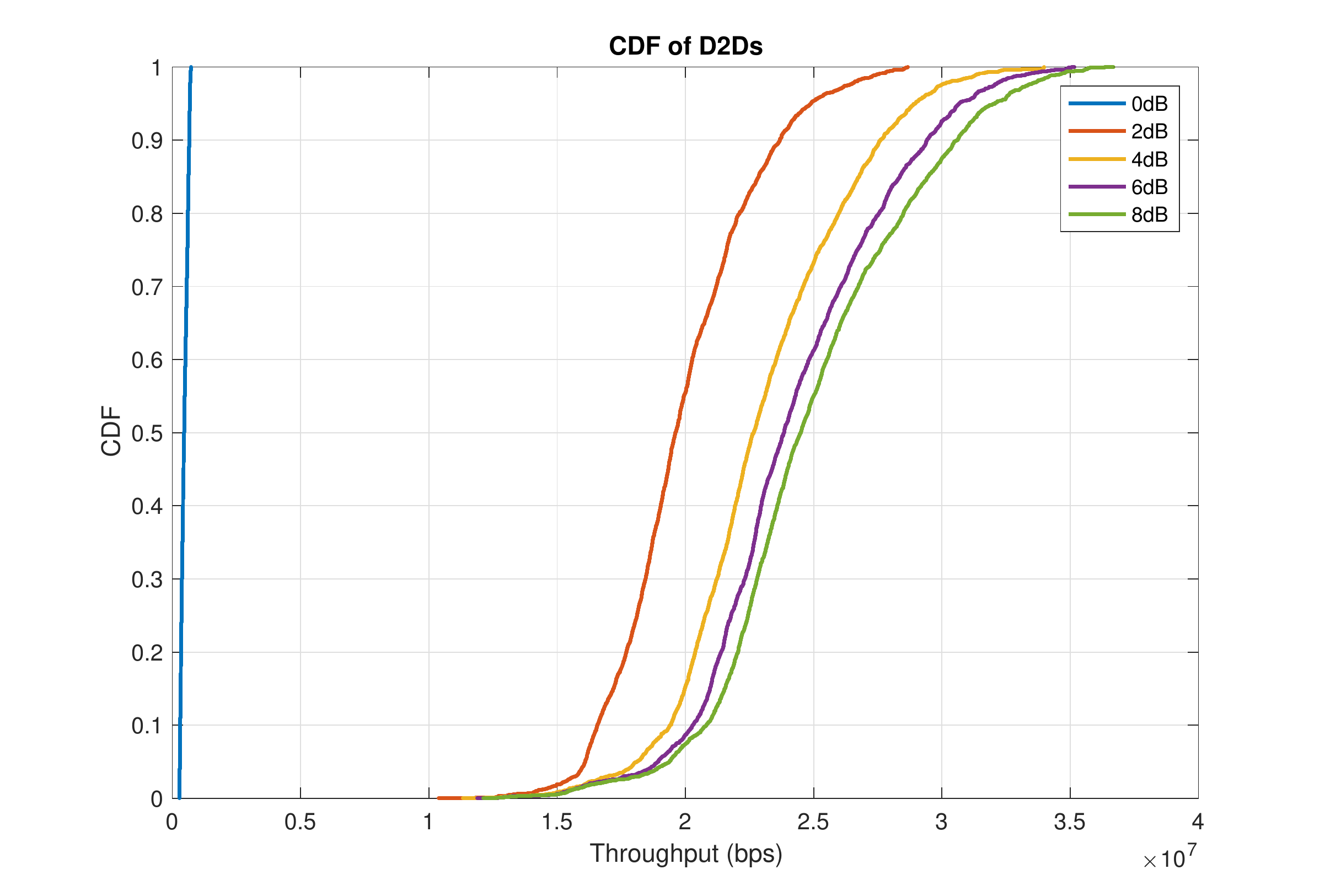}
 \caption{CDF of D2Ds throughput}\label{fp21}
\end{figure}

For a Poisson arrival rate of 20, Fig.~\ref{se10}, \ref{se6} shows the number of D2D links which reused the spectrum at $SINR_{th}$ 8dB and 6dB respectively. 
When the $SINR_{th}$ is minimum (for eg: 6 dB) the algorithm allows less number of D2D links to be initiated in the serving BS to prevent the overall system performance degradation. 
Hence remaining D2D links are blocked (as shown in Fig.~\ref{br6}) as it may create interference to the existing UE in the serving cell. 
When there is a increase in $SINR_{th}$ (\emph{i.e.,} 8 dB) we observe more D2D links being served as shown in Fig. \ref{se10}, and consequently lesser D2D links are blocked as shown in Fig. \ref{br10}. 
Figs. \ref{se10}, \ref{se6}, \ref{br10} and \ref{br6} show how the number of D2D links reusing the spectrum is influenced by utilizing either only the serving cell spectrum or by utilizing both the the serving cell spectrum in conjunction with the neighboring spectrum. 
It can be observed that the D2D links initiated while utilizing the serving cell spectrum alone are not sufficient to satisfy the traffic demands. 
Hence, one can intelligently use the neighboring cell spectrum to enable more D2Ds links.
Fig.~\ref{fp22} shows the variation of UEs throughput as we vary the $SINR_{th}$ according to Algorithm 1. 
This $SINR_{th}$ is the overall permissible degradation of the UEs SINR in each TTI. 
When the degradation is 0 dB (\emph{i.e.,} no drop in original the SINR of the UE is allowed), very few allowable D2D links are initiated and the UEs throughput is maximum aswell. 
As the $SINR_{th}$ increases, number of D2D links which can be served also increases and hence  UEs throughput decreases due to interference from more D2D links as shown in Fig.~\ref{fp21}. 
As the $SINR_{th}$ varies from 0 dB to 8 dB in steps of 2 dB, UEs throughput decreases gradually for 2, 4, 6 and 8 dB with respect to 0dB. It can be noted that, even if we compromise the $SINR_u$ as high as 8 dB, there is no drastic degradation in UEs throughput.   
\par
\begin{figure}
\includegraphics[scale=.36]{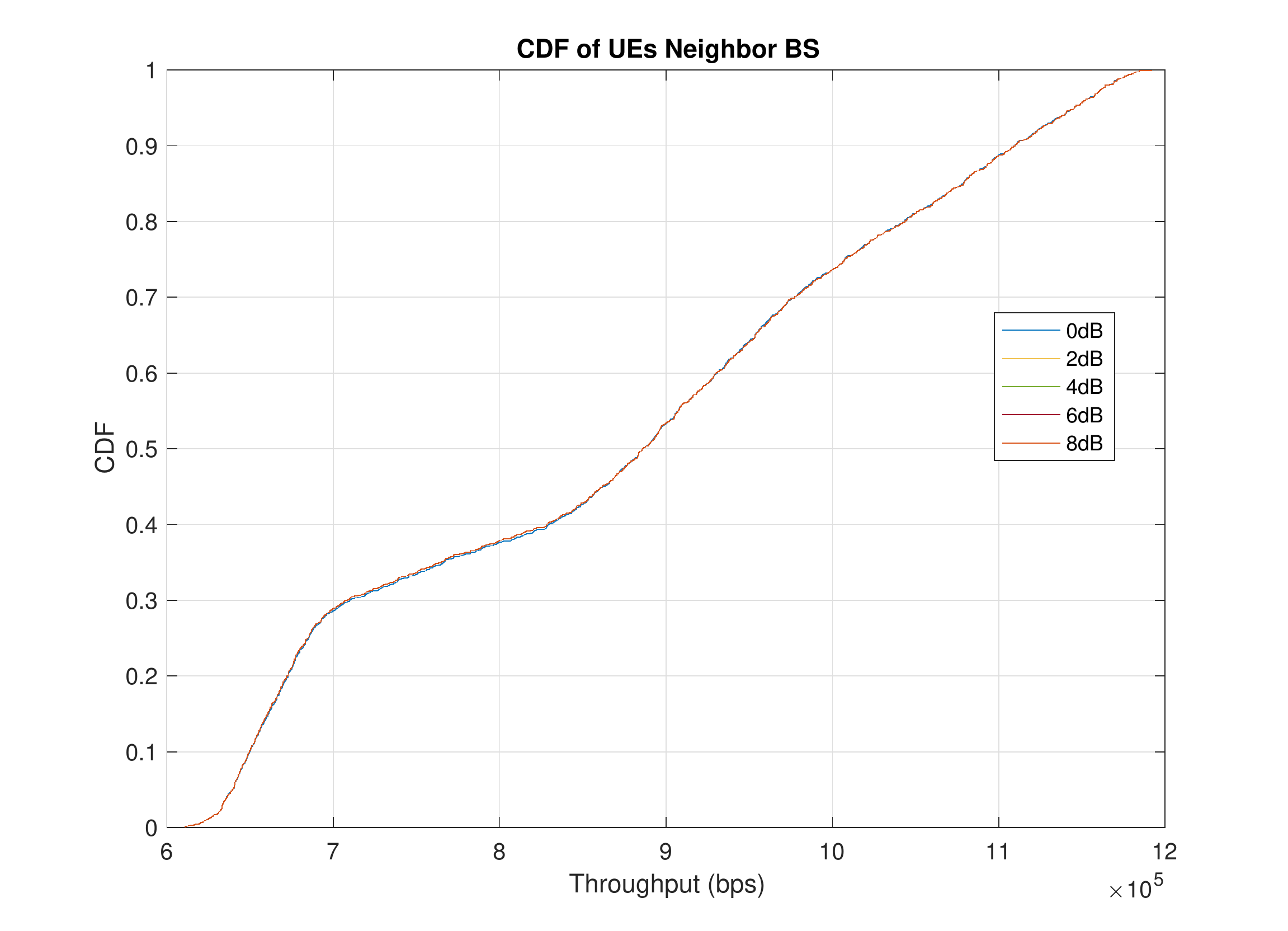}
 \caption{CDF of Neighboring UEs throughput}\label{fp24}
\end{figure}

\begin{figure}
\includegraphics[scale=.36]{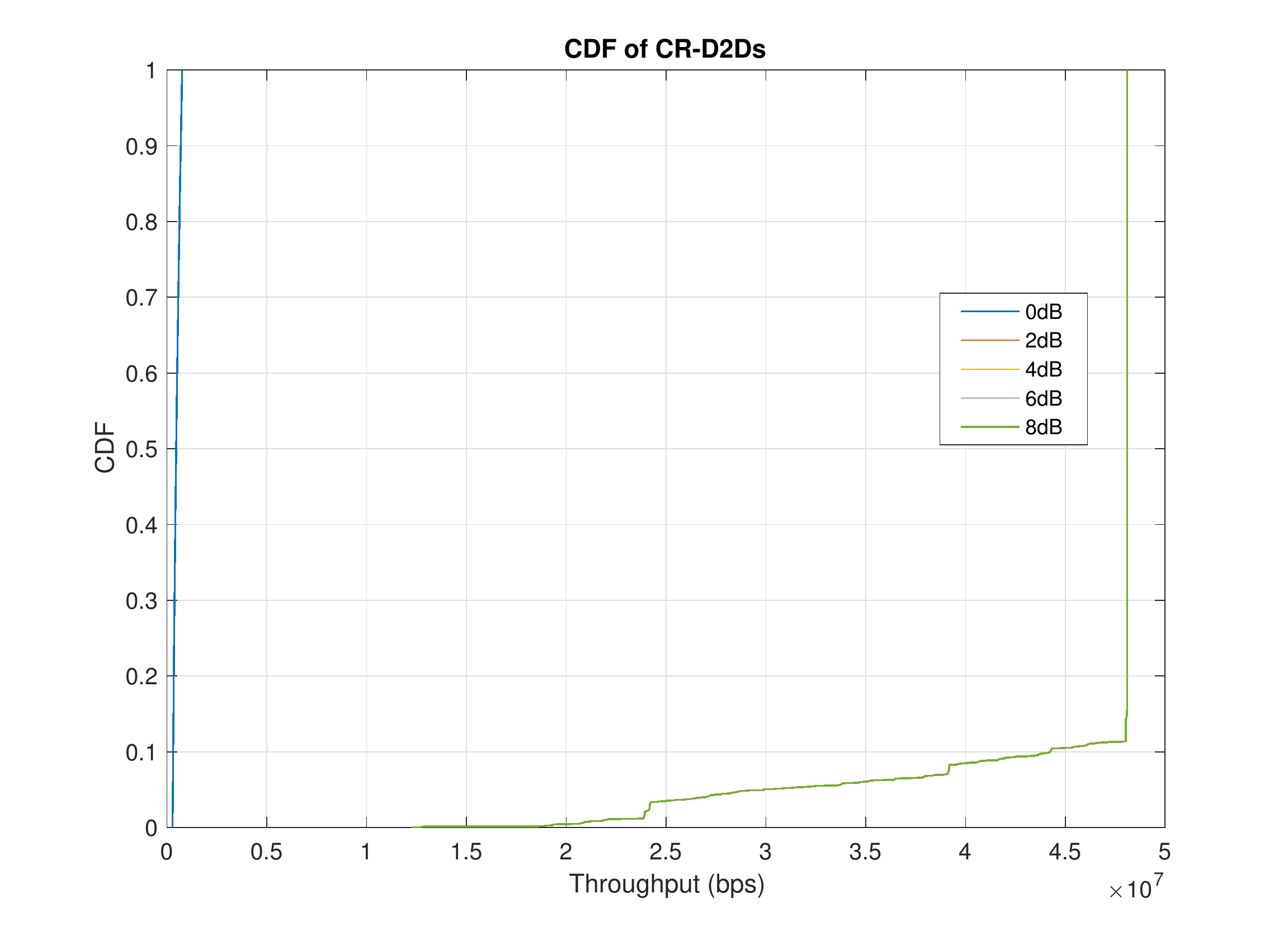}
 \caption{CDF of CR-D2Ds throughput}\label{fp23}
\end{figure}

 As shown in Fig.~\ref{fp21} and \ref{fp23}, an increase in the number of D2D links will increase the overall system performance. 
To visualize the advantage from a cognitive aspect, our LTE-CR-D2D algorithm intelligently reuses the  spectrum (\emph{radio resources}) from the neighboring base station (\emph{i.e.,} inter operator spectrum sharing) and enable more D2D links (as shown in Fig.~\ref{fp23}), while at the same time the spectrum is used by the primary users in the neighboring BS. 
Since the D2D links are faraway from the neighboring base station, it will not impact the neighboring BS throughput (as shown in Fig.~\ref{fp24}).
\section{Conclusion}
In this paper, we contend that cognitive radio based D2D communication has the potential to increase the spectrum efficiency and cellular network capacity. CRD2D intelligently uses all the available spectrum inside the serving cell, by utilizing neighboring BS spectrum to fairly coexist with neighboring LTE UEs. 
Simulation results showed that the spectral reuse and throughput of the overall system increased significantly without compromising on the throughput of the UEs in the serving and neighboring cells. 
We also analyzed the spectrum efficiency of the D2D with and without CR capability. 

\end{document}